\documentstyle[pre,aps,epsfig]{revtex}
\begin{document}

\title{Transition to spatially periodic patterns in nematics under 
oscillatory shear flow: linear analysis}               

\author{O.S. Tarasov$^{\footnotesize{1,2}}$, A.P. Krekhov$^{\footnotesize{1,2}}$ 
and L. Kramer$^{\footnotesize{1}}$}

\address{
$^{1}$Institute of Physics, University of Bayreuth, D-95440 Bayreuth,
Germany}

\address{
$^{2}$Institute of Molecule and Crystal Physics, Russian Academy of
Sciences, 450025 Ufa, Russia}

\date{\today}

\maketitle

\begin{abstract}
We consider the orientational instabilities, both homogeneous and spatially
periodic, developing in a nematic liquid crystal under
rectilinear  oscillatory Couette flow for director 
alignment perpendicular to the flow plane. Using numerical and 
analytical approaches we determine the critical amplitude of oscillatory
flow instabilities, the critical wave number and 
the symmetry of the destabilizing mode. It was found, that by varying of
the oscillatory flow frequency the instability changes its 
temporal symmetry. The mechanism of this transition is coupled with
the inertia of the nematic fluid. We also shown that an electric field 
applied to
the nematic layer can induce switching  between instabilities with
different spatial and temporal symmetries. 
The complete phase diagram of the flow instabilities is presented.    
\end{abstract}
\vspace{24pt}


\section{Introduction}

Nematic liquid crystals (nematics) 
represent one of the simplest anisotropic fluid \cite{deGennes}. 
In order to describe the macroscopic dynamics of these materials 
made up from elongated molecules one introduces as additional hydrodynamic 
variable a unit vector {\bf n}, the {\it director}, which describes  the
axis of local preffered molecular orientation. The strong coupling 
between the director and velocity field provides a number of interesting
phenomena in nematics under flow of different types, such as pattern
formation \cite{edvpmbook,obzor}, director tumbling \cite{zunles}, 
autowaves \cite{obzor} 
{\it etc.}. The flow behavior of a nematic strongly depends on 
the geometry and material parameters, in particular the viscosity coefficients 
$\alpha_1 \dots \alpha_6$ known as Leslie coefficients. Two of 
these coefficients, $\alpha_2$ and $\alpha_3$, appear in the director
equation and play an particular role in hydrodynamic instabilities. 
$\alpha_2$ is supposed to be negative for materials made of rod-like 
molecules, whereas the sign of $\alpha_2$ is not restricted.
If the nematic is initially oriented {\em within the flow plane} 
($=x,z$ plane) then, in materials with  $\alpha_3<0$ under steady
flow (plane Couette or Poiseuille type) the director tends to align at an angle 
$\theta_{fl}=\tan^{-1}\sqrt{\alpha_3/\alpha_2}$ to the
flow direction in the flow  plane ($=x$ direction), which corresponds to
stationary homogeneous solution of the nematodynamic equations \cite{lesadv}. 
Such nematics are called {\it flow-aligning}.  
The solution $\theta_{fl}$ with appropriate sign of the square root 
was shown to be linearly stable in the
case of steady flows \cite{lesadv,pikin,pikchi}. In oscillatory
flows the director tends to oscillate between $\pm\theta_{fl}$. 
With a nonlinear velocity profile, as in Poiseuille flow, this 
solution can lose stability \cite{krekraPRE96,tarkreC99}. 
In the case $\alpha_3>0$ under steady flow one has tumbling with a 
director profile depending strongly on the shear rate \cite{zunles}. 
Nematics with $\alpha_3>0$ are called {\it non-flow-aligning}.  
The picture of orientational
transitions becomes even richer if the director is initially oriented 
{\em perpendicular
to the flow plane} ($y$ direction). This orientation corresponds, for symmetry 
reasons, to a stationary homogeneous solution of dynamic equations of nematics.
With appropriate surface alignment it can lose stability  at a
certain magnitude of the shear rate via a stationary bifurcation  \cite{pg1,pg}. 
In the absence of external fields in steady Couette as well as Poiseuille flows
the director profile then deforms  homogeneously in space. 
In steady Couette flow a magnetic field applied along the initial director 
orientation changes the type of instability from homogeneous to spatially 
periodic at a critical field strength \cite{pg1}. 
In Poiseuille flow the magnetic field does not affect the type of instability 
\cite{pg}, but rolls can be observed above a secondary instability \cite{pg}. 
Classical experiments of Pieranski and Guyon (PG) on oscillatory 
Couette flow \cite{pg1} and recent experiments \cite{mulpeaPRSLA99,ankapJETP96}
show the appearance
of a roll instability with the critical shear rate depending on flow frequency. 
It was shown by PG, that the temporal
symmetry of the instability is influenced by an applied electric field \cite{pg1}. 
A theoretical treatment of the problem
\cite{pg1,dv,mdv,edvpmbook} gives the mechanisms of the instability 
and is in a good agreement with experimental results, including the 
change in symmetry. For oscillatory Poiseuille flow the situation is
similar\cite{mdv1,jpg,mann}. 
In all the theoretical treatments \cite{pg1,dv,mdv,mdv1,mann} the 
inertia of the fluid is neglected, which is reasonable for 
the low frequencies used. In the absence of an electric field 
one then has no changes in the stability scenario as a function of 
the flow frequency.

In the present work we investigate numerically as well as analytically 
orientational instabilities, both homogeneous and spatially periodic, induced 
by plane, oscillatory Couette flow in nematics in the
geometry with the director pre-aligned {\em perpendicular to the flow plane} in a wide range
of the flow frequencies. We look for homogeneous and spatially 
periodic instabilities and  study the influence of an electric
field on flow instabilities for both, negative and positive, dielectric
anisotropy of nematic. Recent experiments \cite{mulpeaPRSLA99,ankapJETP96}
show that high-frequency range could also be investigated experimentally 
as well.

\section{Governing equations and symmetry}

Let us consider a nematic layer filling the space $-d/2<z<d/2$ 
bounded by infinite parallel 
plates, one of which (upper) oscillating harmonically along $x$ with 
amplitude $A$ and circular frequency $\omega=2\pi f$.  
The director is oriented (rigidly anchored) at the boundaries along $y$, i.e.,
perpendicular to the flow plane. An electric field of strength
$E$  can be applied along $z$.

Before any instability develops (sufficiently small amplitude $A$) the
director remains undistorted and the nematic behaves like an isotropic
fluid with viscosity $\eta_3=\alpha_4/2$. For further consideration let us
introduce the dimensionless variables
\begin{eqnarray}
\tilde{z}=z/d, \quad \tilde{t}=t\omega, \quad a=A/d.
\nonumber
\end{eqnarray}
In following we omit the tildes.
Then the {\em subcritical} velocity profile can be found from Navier-Stokes
equation
\begin{eqnarray}
\epsilon_v^{-1} v_{x_{,t}}^0=v_{x_{,zz}}^0 \; , \quad 
v_x^0\Bigl|_{z=1/2}=a \cos t \;, \quad v_x^0\Bigl|_{z=-1/2}=0.
\label{basveleq}
\end{eqnarray}
Here $\epsilon_v=1/(\omega\tau_v)$ with $\tau_v=\rho d^2 /\eta_3$ the 
viscous relaxation time. Thus, our {\em basic} state is
\begin{eqnarray}
{\bf n}^0=(0,1,0), \quad {\bf v}^0=[v_x^0(z,t),0,0]. \label{initstate}
\end{eqnarray}
Below it will
be shown that for the stability problem the inertia of fluid in the 
basic velocity field can
be neglected, so that the solution of (\ref{basveleq}) is simply 
$v_x^0=a(z+1/2)\cos t$.
In order to investigate the stability of (\ref{initstate}) we
linearize the nematodynamic equations \cite{lesadv} around this state 
\begin{eqnarray}
\begin{array}{lcl}
{\bf n}&=&(0,1,0)+(n_x,n_y,n_z), \\ \nonumber
{\bf v}&=&(v_x^0,0,0)+(v_x,v_y,v_z), 
\end{array}
\label{lin}
\end{eqnarray}
where the perturbations $n_i=n_i(y,z,t), v_i=v_i(y,z,t)$, $\{i=x,y,z\}$ are 
small. Here we neglect the $x$ dependence of perturbations assuming the 
wave vector of instability if not zero is perpendicular to the shear direction. 

Eliminating $v_y$ by use of the incompressibility condition
$v_{y_{,y}}=-v_{z_{,z}}$ and the pressure by cross differentiating the 
nematodynamic equations one obtains
\begin{eqnarray}
\begin{array}{lcl}
(\epsilon_v^{-1} \hat{L}_1+\hat{L}_2)v_x&=&-\epsilon_v^{-1} \partial_{yy}v_z a \cos(t) + 
\alpha_2'\partial_{yyyt}n_x+\\ 
\phantom{(\epsilon_v^{-1} \hat{L}_1+\hat{L}_2)v_x}&+&\frac{\alpha_5'-\alpha_2'}{2} a\cos(t) \partial_{yyy} n_z;
\end{array}
\label{vx}
\end{eqnarray}
\begin{eqnarray}
(\epsilon_v^{-1} \hat{L}_3+\hat{L}_4)v_z=\hat{F}_1 n_z + a\cos(t)\hat{F}_2 n_x;
\label{vz}
\end{eqnarray}
\begin{eqnarray}
(\hat{G}_1+\epsilon_d \hat{G}_2)n_x=(1-\lambda)^{-1} [n_z a\cos(t)+v_{x_{,y}}];
\label{nx}
\end{eqnarray}
\begin{eqnarray}
(\hat{G}_3+\epsilon_d \hat{G}_4)n_z=
(1-\lambda)^{-1}[ \lambda n_{x_{,y}} a\cos(t)+(\partial_{yy}-\lambda\partial_{zz})v_z], 
\label{nz}
\end{eqnarray}
where $\hat{L}_i$, $\hat{G}_i$ and $\hat{F}_i$ are linear differential
operators 
\begin{eqnarray}
\begin{array}{rclrcl}
\hat{L}_1&=&\partial_{yyt}, 
&\hat{L}_2&=&-(\eta_2'\partial_y^4+\partial_{zzyy}), \\
\hat{L}_3&=&\partial_{yyt}+\partial_{zzt},
&\hat{L}_4&=&-[\eta_2'\partial_y^4+(\alpha_1'+\eta_1'+\eta_2')\partial_{zzyy}+
\eta_1'\partial_z^4], \\
\hat{F}_1&=&\alpha_2'\partial_{yyyt}-\alpha_3'\partial_{zzyt}, 
&\hat{F}_2&=&\frac{\alpha_6'-\alpha_3'}{2}\partial_y^3-\frac{\alpha_3'+\alpha_6'}{2}\partial_{yzz}, \\
 \hat{G}_1&=&\partial_{t}, 
&\hat{G}_2&=&-(k_3\partial_{yy}+k_2\partial_{zz}), \\
 \hat{G}_3&=&\partial_{yt}, 
&\hat{G}_4&=&-[k_3\partial_{y}^3+k_3\mbox{sgn}(\epsilon_a)\pi^2 E_0^2
\partial_y+\partial_{zzy}],
\end{array}
\nonumber
\end{eqnarray}
and
\begin{eqnarray} 
&\epsilon_d=1/(\tau_d \omega) \; , \quad \tau_d=\gamma_1 d^2/K_{11} \; , 
\quad E_0=E/E_F \; , \nonumber \\
&\alpha_i'=\alpha_i/\eta_3 \; , \quad \eta_i'=\eta_i/\eta_3 \; , \nonumber \\
&\gamma_1=\alpha_3-\alpha_2 \; , \quad \lambda=\alpha_3/\alpha_2 \; ,\nonumber \\
&\eta_1=(\alpha_3+\alpha_4+\alpha_6)/2 \; , \quad
\eta_2=(\alpha_4+\alpha_5-\alpha_2)/2 \; . \nonumber 
\end{eqnarray}
Here  $k_i=K_{ii}/K_{11}$ with $K_{ii}$ the orientation elastic
constants, $\tau_d$ the director relaxation time, 
$E_F=\pi/d\sqrt{K_{11}/(\epsilon_0 |\epsilon_a|)}$ the  Fr\'eedericksz
field, $\epsilon_a$ the anisotropy of the dielectric permeativity.
Notation $f_{,i}\equiv \partial f/\partial i$ has been used throughout. 
We introduce the quantity $P_\rho=\tau_v/\tau_d=\rho K_{11}/(\eta_3\gamma_1)$, 
which is of order $10^{-6}$ for ordinary nematics.

The boundary conditions ({\it fully rigid}) read
\begin{eqnarray}
\begin{array}{ll}
v_x\Bigl|_{z=\pm 1/2}=0, & v_z\Bigl|_{z=\pm 1/2}=v_{z_{,z}}\Bigl|_{z=\pm 1/2}=0, \\ 
n_x\Bigl|_{z=\pm 1/2}=0, & n_z\Bigl|_{z=\pm 1/2}=0. 
\end{array}
\label{bc}
\end{eqnarray}

\section{STABILITY ANALYSIS}
\subsection{General formulation}

Let us collect all perturbations in the vector 
${\bf Y}=(v_x,v_z,n_x,n_z)^{\mbox{T}}$.
From the Floquet theorem one can write the model solutions of 
(\ref{vx}-\ref{nz}) in the form
\begin{eqnarray}
{\bf Y} = e^{\sigma t} e^{iqy} \sum_{k=-\infty}^\infty {\bf y}_k(z) e^{k i t},
\label{galdec} 
\end{eqnarray}
where $\sigma$ is the growth rate, $q$ is a wave number of instability.
Instead of solving the Floquet problem (\ref{galdec}) directly we will 
employ a Galerkin expansion \cite{gal}: 
\begin{eqnarray}
{\bf y}_k &=& \sum_{n=1}^\infty  {\bf c}_{nk} \phi_n(z) \; ,
\label{galexp}
\end{eqnarray}
where ${\bf c}_{nk}$ are constant coefficients and ${\phi_n(z)}$ is a set of
trial functions, which satisfy boundary conditions (\ref{bc}), 
Chebyshev functions \cite{Abramowitz}
\begin{eqnarray}
\phi_n(z)=\left\{  
\begin{array}{ll}
T_{2m}(2 z)  - T_0(2 z), & n=2m-1, \\
T_{2m+1}(2 z)- T_1(2 z), & n=2m,
\end{array} \right.
\end{eqnarray}
for $v_x,n_x$ and $n_z$ and Chadrasekhar functions \cite{chandrabook}
\begin{eqnarray}
\phi_n(z) =\left\{ 
\begin{array}{ll}
\cosh(\lambda_m z)/\cosh(\lambda_m/2)-\cos(\lambda_m z)/\cos(\lambda_m/2), &
n=2m-1, \\ 
\sinh(\nu_m z)/\sinh(\nu_m/2)-\sin(\nu_m
z)/\sin(\nu_m/2), & n=2m 
\end{array} \right.
\end{eqnarray}
for $v_z$. Here $\lambda_m$ and $\nu_m$ are the roots of the appropriate 
characteristic equations \cite{chandrabook}.

Due to the structure of Eqs.(\ref{vx}-\ref{nz}) the model solutions 
(\ref{galdec})-(\ref{galexp}) have 
the definite parity in $z$ and $t$ (parity of Fourier modes). Thus, 
the symmetry types are divided into four groups (Table \ref{tab1}). 
The $z$-parity of $v_y$ is opposite to that of $v_z$ whereas the $t$-parity 
is the same. Type I corresponds to the $Y$-symmetry and type
II to the $Z$-symmetry in the notations of PG \cite{pg1}. Type I and III 
suppose that $v_z$ can have nonzero time averaged and rolls in this case
are {\em overturning}, i.e. they rotate steadily. Rolls of II and IV type 
are {\em oscillating}.

Substituting (\ref{galdec})-(\ref{galexp}) into (\ref{vx}-\ref{nz}) and 
projecting
\cite{gal} onto the Galerkin modes one obtains a linear algebraic system 
for the expansion coefficients  ${\bf c}_{nk}$ of the form
\begin{eqnarray}
({\bf A}+a {\bf B}) {\bf c}=\sigma {\bf C} {\bf c}. 
\label{mateq}
\end{eqnarray}
The expansions (\ref{galdec})-(\ref{galexp}) are then truncated and 
the eigenvalue problem (\ref{mateq}) can be solved. The 
condition $\Re[\sigma(a,q)]=0$ for the largest sigma yields the neutral 
curve $a_0(q)$.
If one assumes a stationary bifurcation, i.e.
$\Im(\sigma)=0$ at the threshold, then one has to solve the
eigenvalue problem 
\begin{eqnarray}
({\bf A}^{-1} {\bf B}) {\bf c} = - \frac{1}{a} {\bf c}, 
\label{matsim}
\end{eqnarray}
which is much more convenient. The critical value $a_c$ is given by
$a_c=\min_q a_0(q)$, which also yields the critical wave number $q_c$.
Using (\ref{mateq}) we have checked that the first bifurcation is 
really always stationary \cite{prim}.

\subsection{Results without electric field}

The calculations were done for material MBBA, see App. \ref{appenmatpar} for
material parameters.

\begin{figure}[h]
\begin{center}
\parbox{13cm}{
\vspace*{3cm}
\epsfxsize=13cm
\epsfysize=13cm
\epsfbox{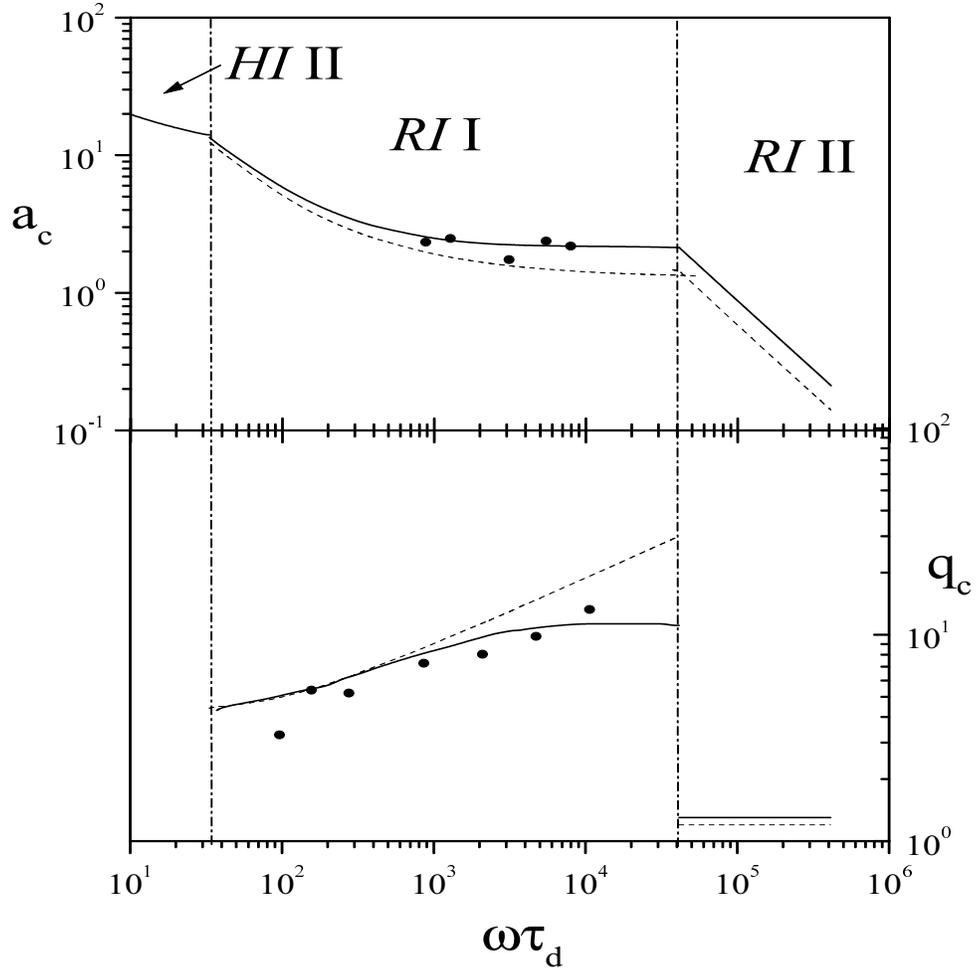}}
\end{center}
\vspace*{0cm}
\caption{Critical amplitude $a_c$ {\em vs.} $\omega\tau_d$ for MBBA. Solid lines
are numerical calculations, dashed lines are approximations
(\protect\ref{acsimp}), points are experimental data from
\protect\cite{ankapJETP96}. {\em RI} denotes roll instability, {\em HI} 
is homogeneous instability, Roman
numbers denote the group of symmetry from table \protect\ref{tab1}.}  
\label{aw2-800}
\end{figure}

For $\omega \tau_d \leq 35$
the first instability developing in the absence of external fields is a
homogeneous one ($q_c=0$) \cite{pg1}. This could be expected already 
from the fact that in steady shear flow the
homogeneous instability occurs first  \cite{pg1}. At
$\omega \tau_d \sim 35$ PG have observed the transition to rolls of 
type I. Up to now the frequency region $\omega \tau_d > 35$ 
remains experimentally poorly investigated. Our analysis for 
$\omega \tau_d \geq 35$  shows that roll instability of
type I persists up to $\omega \tau_d \sim 4.1\cdot 10^4$, where a
transition to the rolls of II-type takes place (Fig.\ref{aw2-800}). At the
transition point the wave number jumps from 12 to 1.2. 
We have checked numerically that the appearance of this transition is related 
to the inertia of the
nematic described by the parameter $\epsilon_v$. By neglecting the inertia
term in the Navier-Stokes Eqs. (\ref{vx}-\ref{vz}), i.e. for 
$\epsilon_v^{-1}\rightarrow 0$, no such transition was found. 
Numerical analysis also shows that thresholds for instabilities of
III and IV types are higher for all frequencies investigated. 

In order to understand the mechanism leading to this transition let us 
analyze Eqs.(\ref{vx})-(\ref{nz}). 
By acting with $\partial_y$ on Eq. (\ref{vx}) and with 
$\partial_{yy}-\lambda\partial_{zz}$ on (\ref{vz}) one may
eliminate the velocities. In the remaining equations for $n_x$ and $n_z$
we use a one-mode spatial approximation
\begin{eqnarray}
\begin{array}{lcl}
n_x&=&\cos(\pi z) \cos(qy) f(t), \\
n_z&=&\cos(\pi z) \cos(qy) g(t),
\end{array}
\nonumber
\end{eqnarray}  
appropriate for the symmetry of type I or II. 
Projecting the equations one obtains coupled second order ODEs 
for $f(t)$ and $g(t)$. Further
simplifications can be made by neglecting the 
terms proportional to $P_{\rho}$ and $\epsilon_v^{-3}$. 
One then arrives at  
\begin{eqnarray}
&&\dot{f}+\epsilon_d m_1 f+( \epsilon_v^{-1} m_3 a \sin\!t
-m_2 a \cos\!t) g=0, \label{nxex} \\
&&\dot{g}+\epsilon_d l_1 g - l_2 a \cos\!t f=0. 
\label{nzex}
\end{eqnarray}
Coefficients $m_i(q), l_i(q)$ are presented in Appendix \ref{appencoeff}. 
In looking for expressions for the critical amplitude we assume
\begin{eqnarray}
\begin{array}{lcllcll}
f&=&f_{osc},       & g&=&g_{st}+g_{osc} &\mbox{ for type I}, \\
f&=&f_{st}+f_{osc},& g&=&g_{osc}         &\mbox{ for type II},   
\end{array} \label{subav}
\end{eqnarray}
that corresponds to the separation of the functions $f$ and $g$ into 
the stationary (index $st$) and oscillatory ($osc$) parts. Substituting
(\ref{subav}) into (\ref{nxex})-(\ref{nzex}) and using time averaging
procedure one obtains the critical amplitudes for roll instabilities of
I and II type
\begin{eqnarray}
\begin{array}{lcl}
a_c^{I}&=&\left[2\frac{(\epsilon_d^2 m_1^2+1)}
{[m_1 m_2 +(\epsilon_v\epsilon_d)^{-1}m_3]}\frac{l_1}{l_2}\right]^{1/2}, \\
a_c^{II}&=&\left[2\frac{(\epsilon_d^2 l_1^2+1)}
{[m_2 l_1 -(\epsilon_v\epsilon_d)^{-1} m_3]}\frac{m_1}{l_2} \right]^{1/2}.
\end{array}
\label{acsimp}
\end{eqnarray}  
Let us analyze the formulas (\ref{acsimp}). From (\ref{coeff_mi}) one has 
that $l_2(q)$ is positive for large $q$ ($>2$)
and very small $q$ ($\ll 1$).
Taking into account that $l_1(q)$ is positive for all $q$ and the 
expression $[m_1 m_2 + m_3 (\epsilon_v\epsilon_d)^{-1}]$ remains positive up to very 
large value of
$\omega\tau_d$ ($\sim 10^6$) one sees from first Eq.\,(\ref{acsimp}) that
in order to have a threshold $l_2$ has to be
positive. Thus, only large $q$ (rolls) and homogeneous instabilities can
develop with symmetry of I type. In the second Eq.(\ref{acsimp}) the expression 
$[m_2 l_1 - m_3 (\epsilon_v\epsilon_d)^{-1}]$  in the denominator changes sign 
with increasing frequency from positive to negative at the frequency
given by  
\begin{eqnarray}
(\omega \tau_d)^2_{tr}=\frac{m_2 l_1}{m_3} \frac{1}{P_{\rho}} \; ,
\label{omtr}
\end{eqnarray}  
A lower limit of the transition frequency is
given by setting $q=0$ in (\ref{omtr}). 
Below $(\omega \tau_d)_{tr}$ one has
$[m_2 l_1 - m_3 (\epsilon_v\epsilon_d)^{-1}]>0$.  Since $m_1>0$ for all $q$ 
one needs $l_2>0$ to have a threshold, i.e., only
the large $q$ instability (rolls) can exist. For
$\omega\tau_d>(\omega\tau_d)_{tr}$ one needs negative $l_2$, which provides 
an instability with small $q$.
One can conclude from (\ref{acsimp}) that only the product
$\epsilon_v\epsilon_d$ plays an essential role in the transition from rolls of 
I type symmetry to rolls of II type, not $\epsilon_v$ itself. 

Let us now return to the question of
neglecting the inertia term in Eq.(\ref{basveleq}) for the basic velocity. 
Assuming $\epsilon_v^{-1}$ to be small 
(for $\omega\tau_d\ll 1/P_{\rho}\sim 10^6$) one can find the expression
for basic velocity $v_x^0$ as an expansion in $\epsilon_v^{-1}$. 
Keeping only the first correction changes the threshold amplitude [and, consequently, 
$(\omega\tau_d)_{tr}$] by less than 1\%, so that $\epsilon_v^{-1}$
can really be neglected in (\ref{basveleq}). 

The results for the approximate expressions (\ref{acsimp}) are included in
Fig.\ref{aw2-800}. The formulas (\ref{acsimp}) have an error
of about $30\%$ at small $\omega\tau_d$, which grows further with $\omega\tau_d$. 
This is an obvious consequence of the one-mode approximation. Nevertheless,
the analytic expressions show all the important features of the 
numerically calculated
thresholds, including the switching between the modes of I and II types. 
However, one
should note that the frequency dependence of the critical wave number $q_c$ in
the one-mode approximation does for I type rolls not show the saturation like 
in the numerical calculations.

For $\omega \tau_d \gg 1/\sqrt{P_{\rho}}$, when $1/(\epsilon_v\epsilon_d) \gg
1$, one can
expand Eqs.\,(\ref{acsimp}) in $1/(\epsilon_v\epsilon_d)$. For the 
threshold of II type rolls one gets   
\begin{eqnarray}
a_c^{II}=\left[-2\frac{(\epsilon_d^2 l_1^2+1)}
{m_2 l_1} \frac{m_1}{l_2} P_{\rho} \right]^{1/2}
\frac{1}{\omega\tau_d}. 
\end{eqnarray} 
Since $\epsilon_d \ll 1$ for $\omega\tau_d \gg 1$ one can see that 
$a_c^{II}$ is inversely proportional to $\omega\tau_d$.   

By increasing $\omega\tau_d$ the critical amplitude $a_c(\omega\tau_d)$ 
of mode I
appears to tend to a constant value already at rather low $\omega\tau_d$, 
where $\epsilon_v^{-1} \ll 1$. Such behavior was found in 
\cite{demec}
by an one-mode analysis. One can show 
that this feature is not a artefact of the one-time-mode approximation. The same 
result can be obtained by inspection of the high-frequency limit in (\ref{nxex})-(\ref{nzex}) 
(see Appendix \ref{appenhf}).

Summarizing, we conclude from our analysis that increasing of
the oscillatory flow frequency (for given cell thickness $d$) leads to
switching from I to II symmetry type
roll instability. It has been shown that this transition is caused by
the inertia of the nematic fluid described by the parameter $\epsilon_v^{-1}$. 
For MBBA material parameters one has the
transition at $\omega\tau_d \sim 4.1\cdot~10^{4}$. Consequently, in the PG
experiment \cite{pg1} with the same liquid crystal and $d=240~\mu$m the transition
frequency should occur of about $7$~Hz (maximal frequency in this experiment
was $\sim$~2 Hz). Recent experiments \cite{mulpeaPRSLA99}, \cite{ankapJETP96} 
show that this transition could be observed.

\subsection{Comparison with experiments}

Experiments of Mullin and Peacock (MP) \cite{mulpeaPRSLA99} performed
with the nematic E7 show the  existence of a roll instability at least up to 
160 Hz with $d$ between $20$ and $50 \mu$m.  MP report, that the observation 
of probe particle have not shown any macroscopic flow, which could mean 
that the rolls 
are of II, i.e., the oscillating type. We have found in the literature 
only the values of the elastic constants for  E7 \cite{ele7}, so that we are
not able to perform a direct comparison of our
theoretical results with the MP experimental data. One can consider the influence 
of material parameters variation on the instability type. The results
of this consideration are presented in table \ref{tab2}.   
A good test for the quality of
oscillatory Couette flow realized in experiments is scaling law,
which says, that $A/d=F(\omega\tau_d^2)$, where $F$ is some function to be determined
from nematodynamic equations. The experimental data of MP
\cite{mulpeaPRSLA99} satisfy this test only in a rather narrow region of low
frequencies,  namely, up to about 40 Hz. At higher frequencies the scaling law
becomes broken, which could be caused by failure of surface anchoring.

Oscillatory shear flow was investigated experimentally also by
Anikeev and Kapustina \cite{ankapJETP96}. The authors used an 
eutectic mixture of
MBBA and EBBA, which is known to have almost the same material parameters as
MBBA. A geometry where the director is anchored in the plane of the layer 
at an angle $\psi$ to the shear direction was investigated. For
the case $\psi=90^o$, which we are interested in, the frequency dependence of
$a_c$ and $q_c$ are presented.  The
experimental points taken from  \cite{ankapJETP96} are given in 
Fig.\ref{aw2-800} together with theoretical curves. The agreement between
experimental and theoretical data is good.

\subsection{Influence of electric field on the instability
type and symmetry}

\begin{figure}[t]
\begin{center}
\parbox{13cm}{
\vspace*{2.5cm}
\epsfxsize=13cm
\epsfysize=13cm
\epsfbox{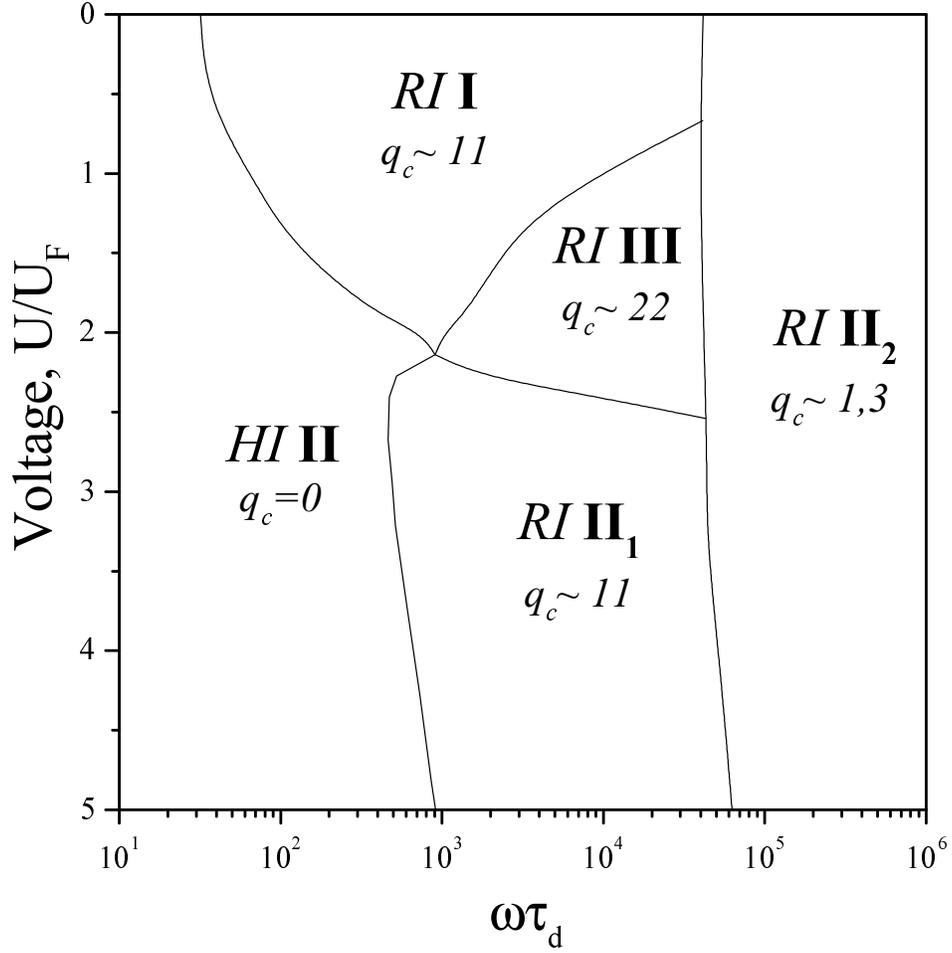}}
\end{center}
\vspace*{-0.5cm}
\caption{Phase diagram of instabilities induced by oscillatory shear flow
and affected by electric field  a nematic with {\em negative} dielectric
anisotropy. {\em RI} denotes roll instability, {\em HI} is homogeneous instability, Roman
numbers denote the group of symmetry from table \protect\ref{tab1}.} 
\label{phased}
\end{figure}

\begin{figure}[t]
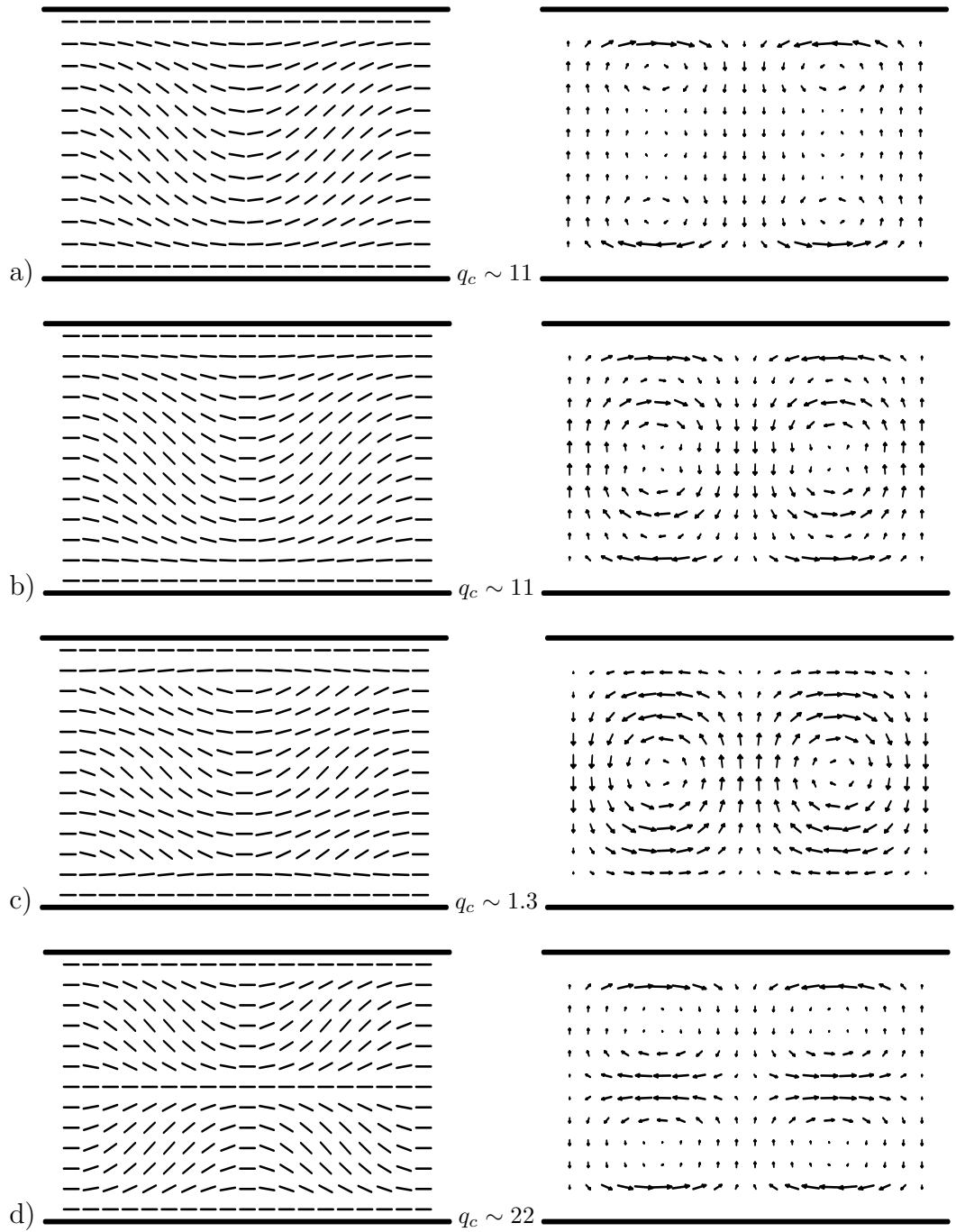

\begin{center}
\parbox{15.5cm}{
{\bf \large a)}
\epsfxsize=6cm
\epsfysize=4cm
\epsfbox{dir1_av.eps} $q_c\sim 11$
\epsfxsize=6cm
\epsfysize=4cm
\epsfbox{vel1_av.eps}} \\
\vspace{0.5cm}
\parbox{15.5cm}{
{\bf \large b)}
\epsfxsize=6cm
\epsfysize=4cm
\epsfbox{dir21_0.eps}  $q_c\sim 11$
\epsfxsize=6cm
\epsfysize=4cm
\epsfbox{vel21_0.eps}} \\
\vspace{0.5cm}
\parbox{15.5cm}{
{\bf \large c)}
\epsfxsize=6cm
\epsfysize=4cm
\epsfbox{dir22_0.eps}  $q_c\sim 1.3$
\epsfxsize=6cm
\epsfysize=4cm
\epsfbox{vel22_0.eps}} \\
\vspace{0.5cm}
\parbox{15.5cm}{
{\bf \large d)}
\epsfxsize=6cm
\epsfysize=4cm
\epsfbox{dir3_0.eps}  $q_c\sim 22$
\epsfxsize=6cm
\epsfysize=4cm
\epsfbox{vel3_0.eps}}
\end{center}
\vspace*{0.5cm}
\caption{Director (left) and velocity (right) profiles: \\
{\bf a)} {\em RI} I, time averaged, \\
{\bf b)} {\em RI} II$_1$, $t = 0$, \\
{\bf c)} {\em RI} II$_2$, $t = 0$,\\
{\bf d)} {\em RI} III, time averaged.}
\label{profiles}
\end{figure}

Applying an electric field across the nematic layer 
can lead to the switching in particular between the different
types of temporal symmetry of rolls induced by oscillatory shear flow. In the
case of MBBA a change of the roll instability of I type to II type by increasing 
the electric field strength was observed \cite{pg1}. We performed the
stability analysis of the linearized  equations (\ref{vx})-(\ref{nz}) in a 
wide
range of oscillatory flow frequency and electric field strength for
MBBA, which has a {\em negative} anisotropy of the dielectric permeativity. We
have found that the electric field can change not only the temporal symmetry of
instability, but also the spatial one. The phase diagram (Fig.~\ref{phased})
presents the area, where the rolls with symmetry type III appear at
threshold. According to their spatial symmetry these rolls represent
a double-layered structure, so that the critical wave number in this area is
about two times larger than otherwise. For low values of
$\omega\tau_d$ an applied electric field leads to an increase
of the area of homogeneous instability ($HI$), so that the electric field suppress the
roll instability. The profiles of the
velocity and the director in the various modes 
are plotted in Fig.~\ref{profiles}.

\begin{figure}[t]
\begin{center}
\parbox{13cm}{
\vspace*{2.5cm}
\epsfxsize=13cm
\epsfysize=13cm
\epsfbox{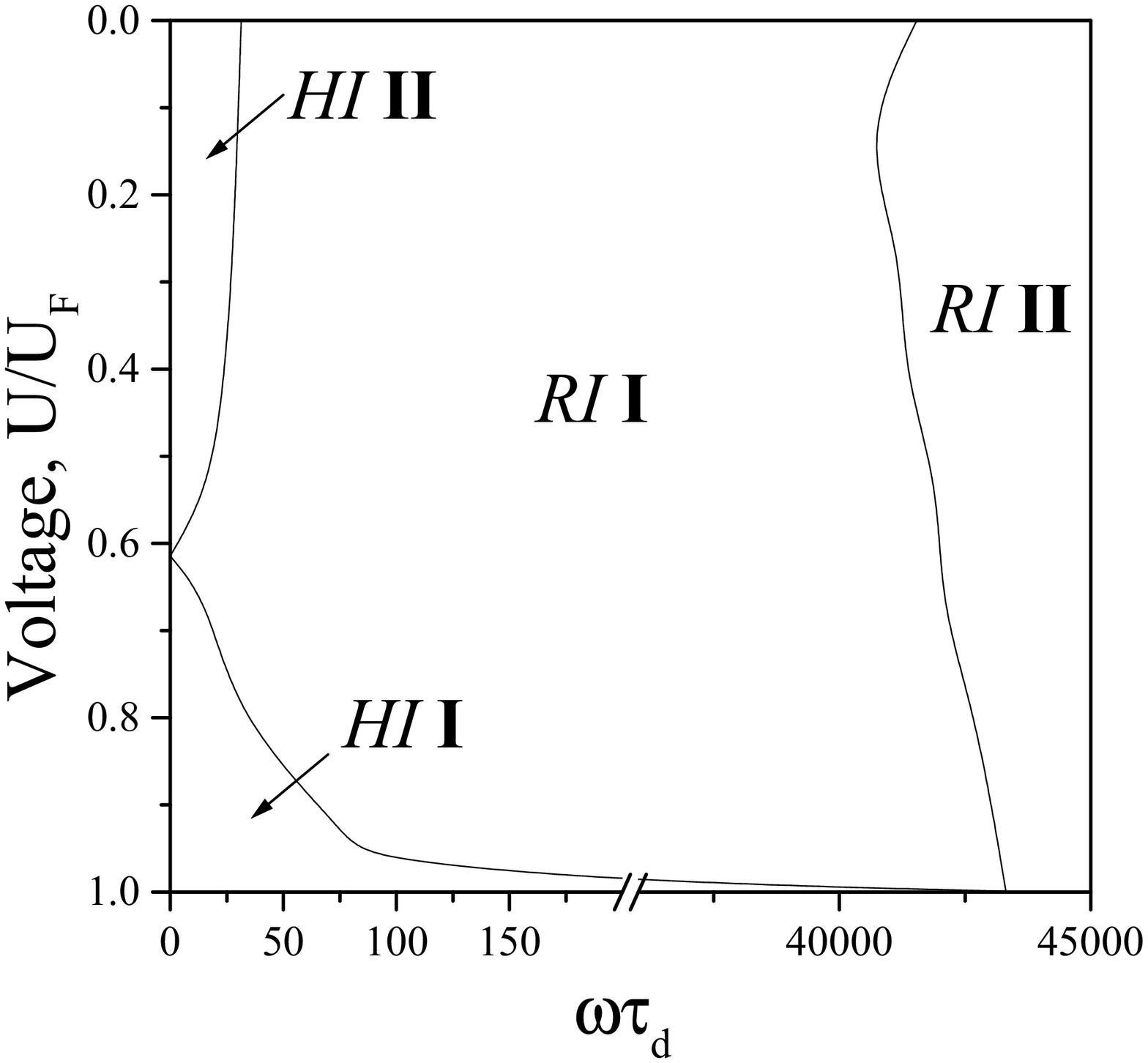}}
\end{center}
\vspace*{-0.5cm}
\caption{Phase diagram of instabilities induced by oscillatory shear flow
and affected by electric field in NLC with {\em positive} dielectric
anisotropy. $U_F$ is Fr\'eedericksz transition voltage. {\em RI} denotes 
roll instability, {\em HI} is homogeneous instability, Roman
numbers denote the group of symmetry from table \protect\ref{tab1}.}  
\label{pdpos}
\end{figure}

We have also calculated the phase
diagram for a nematic with {\em positive} dielectric anisotropy
(Fig.\ref{pdpos}). The material parameters were taken from MBBA except for 
setting the
dielectric anisotropy to be positive. In this case the phase diagram
(Fig.\ref{pdpos}) presents an extensive area of rolls of II symmetry type. 
At low
$\omega\tau_d$ there is a switching between homogeneous instabilities of I 
and II types by varying the electric field. Note, that under these conditions
Fr\'eedericksz transition takes place at voltage $U_F$ ($\sim 3.74$ V). 
As a result, when $U$
approaches $U_F$, the critical shear amplitude $a_c$ goes to zero.

The one-mode analysis in this case gives unsatisfactory agreement with
the full numerical calculations, although it keeps all the important features
(switching between modes {\em etc.}).     

\section{Discussion and conclusion}

We obtained the complete phase diagram of the orientational
instabilities, both homogeneous and spatially periodic, in nematic
liquid crystal subjected to oscillatory Couette flow based on the
linear stability analysis of the full set of nematodynamic equations.
It has been found for the first time that the inertia of nematic fluid,
usually neglected in the theoretical considerations, is responsible
for the new type of roll instability and switching between roll
instabilities with different time-symmetry.
The results of approximate analytical analysis show an essential
features of the influence of inertia of nematic fluid on the
instability scenarios and are in a good agreement with full numerical
simulations of underlying equations.
For the nematic liquid crystal MBBA increasing of flow frequency
leads to the transition between rolls of I and II symmetry types.

Both these modes can be easily distinguished in the experiments.
Although there are no reports concerning the observations of this
transition, the recent experiments on oscillatory flow in nematics
show that the technical possibilities are rich enough to observe
such effect.

We have also investigated the influence of additional electric field
applied across the nematic layer on the flow instabilities for the
liquid crystals with negative and positive dielectric anisotropy.
In case of MBBA it was found that the electric field causes transition
to the double-layered roll structure, which is intermediate regime
between rolls of I and II types observed in \cite{pg1}.
Another interesting point in the experimental investigations could be
the transition between rolls of the same symmetry but quite different
wave numbers at increasing of flow frequency ($RI$II$_1$ and
$RI$II$_2$ in Fig.2).

\section*{Acknowledgments}

\noindent

We wish to thank T. Mullin and T. Peacock for communicating us results
before publication. 
O.T. thanks the Deutscher Akademischer Austauschdienst
(DAAD) for financial support. 
Work supported by Deutsche Forschungsgemeinschaft Grants
No. Kr 690/12, Kr 690/14 and INTAS Grant No. 96-498.

\appendix

\section{Coefficients}
\label{appencoeff}

Coefficients for (\ref{nxex})-(\ref{nzex}) are
\begin{eqnarray}
\begin{array}{lcl}
m_1&=&(\pi^2 k_2+q^2 k_3)(q^2 \eta_2'+\pi^2)/r_1, \\
m_2&=&(1-\lambda)^{-1} [q^2(\eta_2'-\frac{\alpha_5'-\alpha_2'}{2})+\pi^2]/r_1, \\
m_3&=&(1-\lambda)^{-1}/r_1, \\
l_1&=&r_2 [\pi^2 + k_3 (q^2-\mbox{sgn}(\epsilon_a)\pi^2
E^2_0)]/r_3,  \\  
l_2&=&(1-\lambda)^{-1}[\lambda r_2-\frac{\alpha_6'-\alpha_3'}{2}+ \\
\phantom{l_2}&+&q^2\pi^2(\frac{\alpha_6'-\alpha_3'}{2}+
\frac{\alpha_3'+\alpha_6'}{2})-\pi^4 \lambda\frac{\alpha_3'+\alpha_6'}{2}]/r_3, 
\end{array}
\label{coeff_mi}
\end{eqnarray}
with
\begin{eqnarray}
\begin{array}{lcl}
r_1&=&q^2(\eta_2'+(1-\lambda)^{-1}\alpha_2')+\pi^2, \\
r_2&=&q^2[\pi^2(\alpha_1'+\eta_1'+\eta_2')+q^2\eta_2']+\pi^4\eta_1', \\
r_3&=&r_2+(1-\lambda)^{-1}[q^4\alpha_2'-q^2\pi^2(\lambda\alpha_2'+\alpha_3')
+\pi^4\lambda\alpha_3'].
\end{array}\nonumber
\end{eqnarray}

\section{High-frequency limit}
\label{appenhf}

To investigate
the behavior of the critical amplitude in the high-frequency range we 
neglect $\epsilon_v^{-1}$ in
(\ref{nxex})  and (\ref{nzex}) and expand the functions $f$ and $g$ in small
parameter  $\epsilon_d=1/(\omega\tau_d)$. Formally this analysis is valid for
$1 \ll \omega\tau_d \ll 1/P_{\rho}$ (for
MBBA material parameters and $d=20$ $\mu$m this corresponds to $0.02$ Hz $\ll f
\ll 10^6$ Hz).

Setting $\epsilon_v^{-1}=0$ in (\ref{nxex})-(\ref{nzex}) one has in matrix
form 
\begin{eqnarray}
\frac{d}{d t}{\bf Y}={\bf A}(t){\bf Y},
\label{syst_inf} 
\end{eqnarray}
where
\begin{eqnarray}
{\bf Y}=\left( \begin{array}{c}
f\\
g
\end{array} \right), \quad
{\bf A}=\left( \begin{array}{cc}
-\epsilon_d m_1 & -m_2 a \cos t \\
-l_2 a \cos t & -\epsilon_d l_1 
\end{array} \right).
\nonumber
\end{eqnarray}
We seek periodic solution of (\ref{syst_inf}) of the form 
\begin{eqnarray}
{\bf Y}={\bf Y}_0+\epsilon_d {\bf Y}_1+\dots
\nonumber
\end{eqnarray}
At order in $\epsilon_d^0$ one has
\begin{eqnarray}
\frac{d}{d t} {\bf Y}_0 = {\bf A}_0(t) {\bf Y}_0
\label{syst_inf0} 
\end{eqnarray}
where ${\bf A}_0={\bf A}\Bigl|_{\epsilon_d=0}$.
Since ${\bf A}_0(t)$ commutes with ${\bf A}_0(t')$ (unlike ${\bf A}$) the solution
of  (\ref{syst_inf0}) has a form 
\begin{eqnarray}
{\bf Y}_0= e^{\int_0^t d\tau {\bf A}_0(\tau) } {\bf Y}^0, \quad
{\bf Y}^0=\left(
\begin{array}{l}
f^0 \\
g^0
\end{array} \right), 
\label{sol0}
\end{eqnarray}
where $f^0$ and $g^0$ are initial values (fluctuations) of the functions $f_0$ and 
$g_0$, respectively. It can be easily seen, that ${\bf Y}_0$ is periodic 
with time averages $\langle f_0\rangle_t \sim f^0$,
$\langle g_0\rangle_t \sim g^0$, so that the state with ($f^0=0,$ $g^0\ne 0$) 
corresponds to
the mode of I symmetry type,  the state ($f^0\ne 0,$ $g^0=0$) 
corresponds to mode of II symmetry type.
At
first order in $\epsilon_d$ one has the system 
\begin{eqnarray}
\frac{d {\bf Y}_1}{d t} = {\bf A}_0(t) {\bf Y}_1 + {\bf b},
\quad \mbox{with} \quad
{\bf b}= \left( \begin{array}{l}
-m_1 f_0 \\
-l_1 g_0
\end{array} \right) 
\label{syst_mat1}
\end{eqnarray}
The solution of (\ref{syst_mat1}) is given by
\begin{eqnarray}
{\bf Y}_1={\bf Y}_0+
e^{\int_0^t d\tau {\bf A}_0(\tau) }
\int_0^t d\tau e^{-\int_0^\tau d\xi {\bf A}_0(\xi)} {\bf b}(\tau). 
\label{solform}
\end{eqnarray}
Substituting expressions for ${\bf b}$ into 
(\ref{solform}) and using the relation \cite{Abramowitz} 
\begin{eqnarray}
e^{z sin \theta}=I_0(z)+
         2 \sum_{k=0}^\infty (-)^k I_{2k+1}(z) \sin (2k+1)\theta+\nonumber \\
         2 \sum_{k=0}^\infty (-)^k I_{2k}(z) \cos 2k\theta, \nonumber
\end{eqnarray}
where $I_i$ are modified Bessel functions of the first kind,
one has the periodicity condition of ${\bf Y}_1$ 
\begin{eqnarray}
&I_0 (2 \sqrt{m_2 l_2}a)= -\frac{l_1+m_1}{l_1-m_1} 
\quad &\mbox{for I type}, \label{ahfy} \\
&I_0 (2 \sqrt{m_2 l_2}a)= \frac{l_1+m_1}{l_1-m_1} \quad 
&\mbox{for II type}. \label{ahfz}
\end{eqnarray}
Eqs.\,(\ref{ahfy}) and (\ref{ahfz}) represent the connection 
between $a$
and $q$ in implicit form, i.e. the neutral curve $a_0(q)$. 
Minimizing $a_0(q)$ gives the critical values of the flow amplitudes 
$a_c$ and the wave number $q_c$. 
Equation (\ref{ahfz}) has no real roots for $q>0.5$, 
so  
that when the fluid inertia term is neglected only a roll instability
of I symmetry type or homogeneous instability of II symmetry type 
can appear. Eq.\,(\ref{ahfy}) gives a critical amplitude $a_c\sim 2.2$ and 
wave number $q_c \sim 12 $ and for , that 
is in a good agreement with numerical results (Fig.\ref{aw2-800}). From
(\ref{ahfz}) it follows that the smallest threshold $a_c=10.9$ 
corresponds to $q_c=0$.   

The above analysis shows the
role of the elasticity described by terms 
proportional to $\epsilon_d$ in the nematodynamic equations.  
The nature of these terms is singular, i.e. when
neglecting these terms the critical flow amplitude vanishes ($a_c=0$),
although for infinitesimal $\epsilon_d$ there exists a finite value of $a_c$.
Another point is that at high enough frequencies $a_c$ does not depend on 
the flow frequency, which is in contrast to the case of
director alignment in the flow plane \cite{obzor}, where the critical 
amplitude of
roll formation under oscillatory shear flow decreases with  increasing flow
frequency.  

\section{Material parameters}
\label{appenmatpar}

\noindent
The following values of
the MBBA material parameters at 25 $^o$C \cite{je,kn}:

$K_{11}=6.66, K_{22}=4.2, K_{33}=8.61$ (in units $10^{-12}$ N);

$\alpha_1=-18.1, \alpha_2=-110.4, \alpha_3=-1.1,
\alpha_4=82.6, \alpha_5=77.9, \alpha_6=-33.6$
(in units $10^{-3}$ N$\cdot$s/m$^2$).

$\epsilon_a=-0.53$. 

Fr\'eedericksz transition voltage is defined by
$U_F=\pi\sqrt{\frac{K_{11}}{\epsilon_0|\epsilon_a|}}$.


\setlength{\baselineskip}{12pt}

\begin{center}
\begin{table}
\begin{tabular}{c||c|c||c|c||c|c||c|c|}
      & \multicolumn{2}{c||}{I} & \multicolumn{2}{c||}{II} &
       \multicolumn{2}{c||}{III} & \multicolumn{2}{c|}{IV} \\ \hline 
      &  $z$   &  $t$  & $z$ & $t$ & $z$ & $t$ & $z$ & $t$ \\ \hline
$v_x$ & even & odd  & even & even & odd & odd  & odd & even \\
$v_z$ & even & even & even & odd  & odd & even & odd & odd  \\
$n_x$ & even & odd  & even & even & odd & odd  & odd & even \\
$n_z$ & even & even & even & odd  & odd & even & odd & odd 
\end{tabular}
\caption{Spatial ($z$) and temporal ($t$) symmetries of solutions of linearized equations 
(\protect\ref{vx})-(\protect\ref{nz}).}
\label{tab1}
\end{table}
\end{center}  

\begin{center}
\begin{table}
\begin{tabular}{c||c|c|c|c||}
&    Standard values &  Transition values & New $a_c$ & New $q_c$ \\
\hline
$\alpha_1/\alpha_2 $ &  0.164 &  1.01  & 2.38 & 6.9  \\
$\alpha_4/\alpha_2 $ & -0.748 &  -5.92 & 7.72 & 10.1 \\
$\alpha_5/\alpha_2 $ & -0.706 &  -1.83 & 6.64 & 1.4  \\
$\alpha_6/\alpha_2 $ & 0.304  &  -0.85 & 6.60 & 1.4  \\
$K_{11}/K_{22}$      & 1.586  &  5.95  & 2.93 & 7.3  \\
$K_{33}/K_{22}$      & 2.05   &  0.15  & 1.66 & 16.6 
\end{tabular}
\caption{In the second column the normalized standard values for the material parameters 
of MBBA at $25^{o}C$ are given. With those values and at $\omega\tau_d=825$ (typical value 
for experiment \protect\cite{mulpeaPRSLA99}) the roll instability of I type (overturning 
rolls) takes place with critical amplitude $a_c=2.6$ and critical wave number $q_c=7$. When 
any one of the parameters is varied alone a transition to II type rolls (oscillating ones) 
occurs at the value in the third column. In the fourth and fifth columns the new values of 
the threshold amplitude and the critical wave number are given.}
\label{tab2}
\end{table}
\end{center}

\pagebreak

\pagebreak

\pagebreak

\pagebreak

\begin{thebibliography}{12}
\bibitem{prim}
Indeed, there is known numerical effect: periodical force provides
the exictence of parasitic eigenvalues in (\ref{mateq}), which have
non-zero real parts at the threshold amplitude corresponding to the
stationary bifurcation. The imaginary parts of these eigenvalues
are integer numbers, so that
the parasitic eigenvalues are easily to be found and eliminated.
\end{thebibliography}
\end{document}